\documentclass[preprint2]{aastex}

\usepackage{natbib}
\usepackage{graphicx}

\shorttitle{Dissipative damping of the slow magnetoacoustic mode}
\shortauthors{M.S. Marsh and R.W. Walsh}

\begin{document}

\title{Observed damping of the slow magnetoacoustic mode}

\author{M. S. Marsh\altaffilmark{1}, I. De Moortel\altaffilmark{2}, and R. W. Walsh\altaffilmark{1}}
\affil{\altaffilmark{1}Jeremiah Horrocks Institute for Astrophysics \& Supercomputing, University of Central Lancashire, Preston, PR1
2HE, UK}
\affil{\altaffilmark{2}School of Mathematics and Statistics, University of St Andrews, St Andrews KY16 9SS, UK}

\email{mike.s.marsh@gmail.com, mmarsh@uclan.ac.uk}

\begin{abstract}
Spectroscopic and stereoscopic imaging observations of slow magnetoacoustic wave propagation within a coronal loop are investigated to determine the decay length scale of the slow magnetoacoustic mode in three dimensions and the density profile within the loop system. The slow wave is found to have an e-folding decay length scale of $20,000^{+4000}_{-3000}$km with a uniform density profile along the loop base. These observations place quantitive constraints on the modelling of wave propagation within coronal loops. Theoretical forward modelling suggests that magnetic field line divergence is the dominant damping factor and thermal conduction is insufficient, given the observed parameters of the coronal loop temperature, density and wave mode period. 
\end{abstract}


\section{Introduction}\label{sect_intro}
There has been much theoretical work on the propagation of magnetohydrodynamic (MHD) waves in solar coronal plasmas, and the advent of regular space-based observations of the solar corona has subjected theory to empirical test. The theory of MHD waves in magnetic cylinders was described in \cite{edw83}. Since then there have been many papers refining the theory \citep[see reviews by][]{rob00, nak05}. 

There have been many theoretical studies examining the propagation of the slow magnetoacoustic mode in coronal loop structures \citep{por94, nak00, tsi01, ver01, kli04, rob06}. 
Theoretical studies investigating the damping of the slow mode have concentrated on the effects of thermal conduction, compressive viscosity, optically thin radiation, gravitational stratification, magnetic field divergence and shocks \citep{ofm00, ofm02b, dem&hoo03, dem&hoo04, dem04, men04, erd08, ver08}. 

In general, thermal conduction is found to be the dominant mechanism for dissipation of slow modes in the corona. There have been a number of observations of propagation in coronal structures that have been interpreted as manifestations of the slow mode. \cite{ofm97, def98, ofm99} describe propagating intensity disturbances in polar plumes which are found to be consistent with the theory of propagating slow modes. There have also been many observations of coronal loops that are consistent with wave-guiding of the slow mode along the loop structure. These are observed as low amplitude intensity and velocity oscillations located at the base of quiescent coronal loop systems  \citep[see review by][and references within]{dem09} and additionally \cite{me08, wan09}. Recently, \cite{mar10} describe long period oscillations around 10 minutes observed within active regions using Hinode, and \cite{wan11} conducts a review of standing slow modes observed in hot coronal loops. 

The idea of utilizing observations of wave propagation to determine properties of the coronal plasma was first suggested by \cite{uch70} and \cite{rob84}, termed coronal seismology. This idea has since been applied to try and determine coronal properties including damping mechanisms \citep{ofm02} and dissipative coefficients \citep{nak99}.
In \cite{me09a}, observations using the Solar Terrestrial Relations Observatoryt (STEREO) spacecraft were used to measure the phase speed of a slow mode in the corona directly for the first time. This measurement allowed the plasma temperature to be inferred `seismologically' via the propagation speed of the wave through the coronal loop plasma. \cite{me09b} tested these results using spectroscopic observations of the coronal loop system from Hinode; a measurement of the loop temperature and its spatial profile, provided an independent confirmation of the `seismological' results. 

There is a current, and ongoing, debate regarding the nature of propagating disturbances in coronal loop regions as being the signature of flows or waves \citep[see][]{dep10, ver10}. However, it isn't clear if the interpretations in each case are investigating the same observational phenomenon. Undoubtedly, both waves and flows exist in coronal structures, but the question of which is the cause of specific observations of intensity propagations can be difficult to answer. The observations presented in \cite{me09a, me09b} are interpreted as slow magnetoacoustic waves due to the clear periodic nature and the correspondence of the propagation speed to the sound speed at the temperature measured within the loop. 

In this paper, we determine the density profile of the loop system using Hinode observations and measure the amplitude attenuation of the wave as observed from both STEREO spacecraft. This allows the true damping scale length of the slow mode wave to be directly measured in three dimensions for the first time.  This has significant theoretical consequences as it places quantitive constraints on the modelling of slow mode propagation within coronal loops and the dissipation mechanism of the wave.

\begin{figure*}[t]
\centering
\plotone{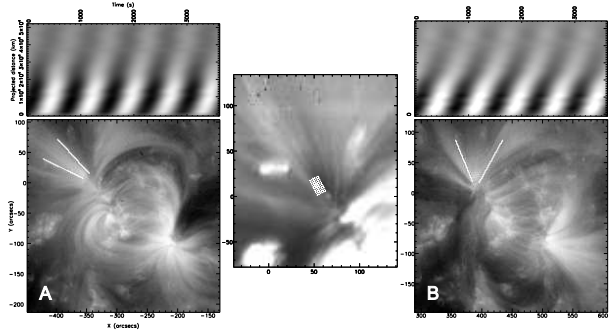}
\caption{Left panel: EUVI 171\mbox{\AA} context image of the coronal loop system analyzed in \cite{me09a}, viewed from STEREO A. The time distance image used to measure the amplitude of the wave propagation is extracted along the path indicated by the white lines. To highlight the low frequency propagation, the time distance images shown are detrended and convolved in the frequency domain with a narrow Gaussian of width 0.14~mHz centred on 1.4~mHz. Center: \ion{Fe}{12} 195\mbox{\AA} intensity image from Hinode/EIS. The loop intensity is extracted from the loop system along the indicated path. The iron emission line ratios are formed using the mean intensity along the path to determine the electron density profile as a function of distance. Right panel: As left panel, but viewed from STEREO B. Note the spatial coordinates for each image are the solar coordinates in the plane of the sky relative to each spacecraft}.
\label{fig1}
\end{figure*}

\section{Observations}
As described in \cite{me09a}, the observations were conducted on 2008 January 10, as part of the Joint Observing Program (JOP) 200 - `Multi-point, High Cadence EUV Observations of the Dynamic Solar Corona'. The STEREO data consists of high (30 second) cadence images in the $171$\mbox{\AA} bandpass from the Extreme-Ultraviolet Imager \citep[EUVI,][]{wue04}. The observations begin at 12:00~UT and observe an active region loop system which supports slow mode wave propagation. 
The left and right panels of Figure~\ref{fig1} show context images of the loop system, along with time distance images (top) extracted along the indicated paths.
See \cite{me09a}, for detailed information on the STEREO/EUVI observations. 

The same coronal loop system is also observed using the Extreme-Ultraviolet Imaging Spectrometer \citep[EIS,][]{cul07} on the Hinode satellite. 
Figure~\ref{fig1}, center panel, shows an EIS raster of the loop system in \ion{Fe}{12}  195\mbox{\AA} line intensity. The spectroscopic raster analyzed here uses the $2$\arcsec slit to build up a $180$\arcsec $\times$ $512$\arcsec rastered image beginning at 18:07~UT. This study contains 24 spectral windows with a selection of lines designed to study active regions. 

\section{Analysis}
\subsection{Preparation of the data}
\subsubsection{STEREO/EUVI}
The data are corrected for detector bias, flat field, and photometric calibration, using the SECCHI\_PREP routines available within the \emph{Solarsoft} database. The pointing of the EUVI images is also co-aligned using the procedure described in \cite{me09a}. 
\subsubsection{Hinode/EIS}
The EIS data are calibrated using EIS\_PREP within {\it Solarsoft}, with standard corrections for dark current, cosmic rays, hot/warm pixels, dusty pixels and an absolute calibration is applied to obtain the data in units of ergs/cm$^{2}$/s/sr/\AA

The pixels along the $y$ axis are binned to increase the signal to noise ratio within the data, resulting in $2\arcsec \times 2\arcsec$ pixels.
The emission line profiles are fitted with multiple Gaussians using the {\it Solarsoft} routine EIS\_AUTO\_FIT\_GEN, and the effects on the line centroids due to the tilt of the EIS slit and the orbital variation are corrected.

\subsection{Loop density profile}\label{loop_density_analysis}
The density profile of the loop system supporting the slow mode propagation is measured using the Hinode/EIS data. The EIS study contains a number of spectral windows that include various iron emission lines. \cite{you09} detail density measurements using \ion{Fe}{12} and \ion{Fe}{13} lines that are observed with EIS. Considering the EIS study analyzed here, the \ion{Fe}{12} (186.88/195.12)\mbox{\AA} ratio may be used to determine the density profile as a function of distance along the loop system, as indicated by the path in the center panel of Figure~\ref{fig1}. The density is calculated at a constant temperature of $0.8$~MK, consistent with the results of \cite{me09b}, using the CHIANTI v6.0.1 atomic database \citep{der97, der09}. To reduce the contamination effect of emission from background plasma, a constant background is subtracted from the data using the intensity of pixel [14,22], which is located in a minimal emission region adjacent to the loop system. 

\subsubsection{The \ion{Fe}{12} (186.88/195.12)\mbox{\AA} ratio}
Before determining the \ion{Fe}{12} density ratio, the fit to the spectral lines, and possible contributions to the line profiles, must be considered carefully. 
As described by \cite{you09}, \ion{Fe}{12} 186.88\mbox{\AA} has a known blend with \ion{S}{11} 186.839\mbox{\AA}. The contribution of this line can be measured via branching ratios, using the \ion{S}{11} 188.68\mbox{\AA} line that appears within the \ion{Fe}{11} 188\mbox{\AA} spectral window, as the two sulphur lines have the fixed ratio of 0.34. 
In the data analyzed by \cite{you09}, the contribution of \ion{S}{11} 186.839\mbox{\AA} to \ion{Fe}{12} 186.88\mbox{\AA} is on the order of 5\% and was not corrected.
In the current data, it is not possible to fit the \ion{S}{11} 188.68\mbox{\AA} line above the noise, indicating that the contribution of \ion{S}{11} 186.839\mbox{\AA} to the \ion{Fe}{12} 186.88\mbox{\AA} line is even more negligible, and thus \ion{Fe}{12} 186.88\mbox{\AA} is fitted with a single gaussian. Additionally, the weak line at 186.98\mbox{\AA}, in the wing of \ion{Fe}{12} 186.88\mbox{\AA}, is negligible and so neglected. 

The \ion{Fe}{12} 195.12\mbox{\AA} line includes a blend with the \ion{Fe}{12} 195.18\mbox{\AA} transition. As mentioned by \cite{you09}, the relative strength of this line is reduced at lower densities, which is consistent with the type of quiescent loops observed here. The \ion{Fe}{12} 195.18\mbox{\AA} line is not found to be significant in this loop system, thus the \ion{Fe}{12} 195.12\mbox{\AA} line is fit with a single Gaussian. 

\subsubsection{Checking the density profile with \\(186.88/192.39)\mbox{\AA} and (186.88/193.51)\mbox{\AA}}
To confirm the density profile results using the \ion{Fe}{12} (186.88/195.12)\mbox{\AA} ratio, the 195.12\mbox{\AA} line may be substituted in the ratio by the density insensitive  \ion{Fe}{12} 192.39\mbox{\AA}, or 193.51\mbox{\AA} lines. These lines are available in the EIS study, within the \ion{Ca}{17} 192\mbox{\AA} window. The (186.88/192.39)\mbox{\AA} and (186.88/193.51)\mbox{\AA} density profiles may then be compared to confirm the density profile obtained using the 195.12\mbox{\AA} line.


\subsection{Fitting the amplitude decay}
A detailed description of the time-distance reduction method and analysis of the wave propagation is given in \cite{me09a}. The amplitude of the wave oscillation is fitted at each spatial location in time distance images (see Figure~\ref{fig1}), from the A and B spacecraft, using the Bayesian fitting code described in \cite{me08}.
This determines the frequency, phase, and amplitude of the wave as a function of distance along the loop system, in the plane of the sky as viewed from each spacecraft. Beginning with the initial point of decay in the amplitude profile, the measured amplitude decay is fitted with a function of the form 
$$A(s)=\beta e^{-\alpha x} + \gamma,$$
where $A(s)$ is the spatial profile of the wave amplitude along the loop system, $\beta$ and $\gamma$ are appropriate constants, $\alpha$ is the decay constant, and therefore $1/\alpha$ is the amplitude decay scale length of the wave in the plane of the sky. 

\section{Results}

\subsection{Loop density profile}
\begin{figure}[h]
\epsscale{1.0}
\centering
\plotone{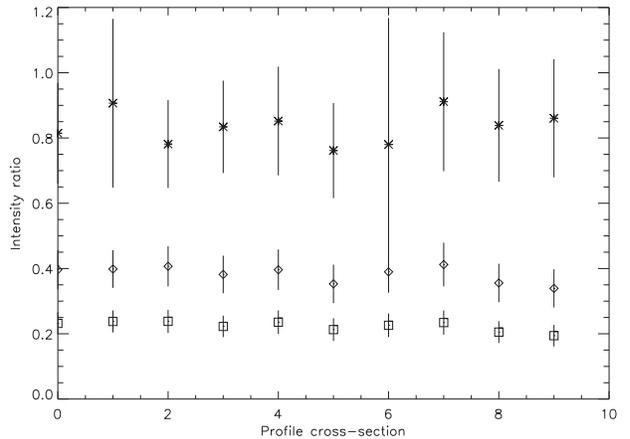}
\caption{Intensity ratio profiles, along the loop path indicated in the center panel of Figure~\ref{fig1}, denoted by (186.88/195.12)\mbox{\AA} (square), (186.88/193.51)\mbox{\AA} (diamond), and (186.88/192.39)\mbox{\AA} (star).}
\label{fig2}
\end{figure}

Figure~\ref{fig2} shows the intensity ratio profiles along the path within the loop system as indicated in the center panel of  Figure~\ref{fig1}. These profiles show the \ion{Fe}{12} (186.88/195.12)\mbox{\AA}, (186.88/193.51)\mbox{\AA}, and (186.88/192.39)\mbox{\AA} ratios. The comparatively smaller errors of the (186.88/195.12)\mbox{\AA} ratio are due to the greater intensity of the 195.12\mbox{\AA} line. The large error in the (186.88/192.39)\mbox{\AA} ratio at cross-section 6 is due to a badly fitted pixel. 

\begin{figure}[h]
\epsscale{0.9}
\centering
\rotate
\plotone{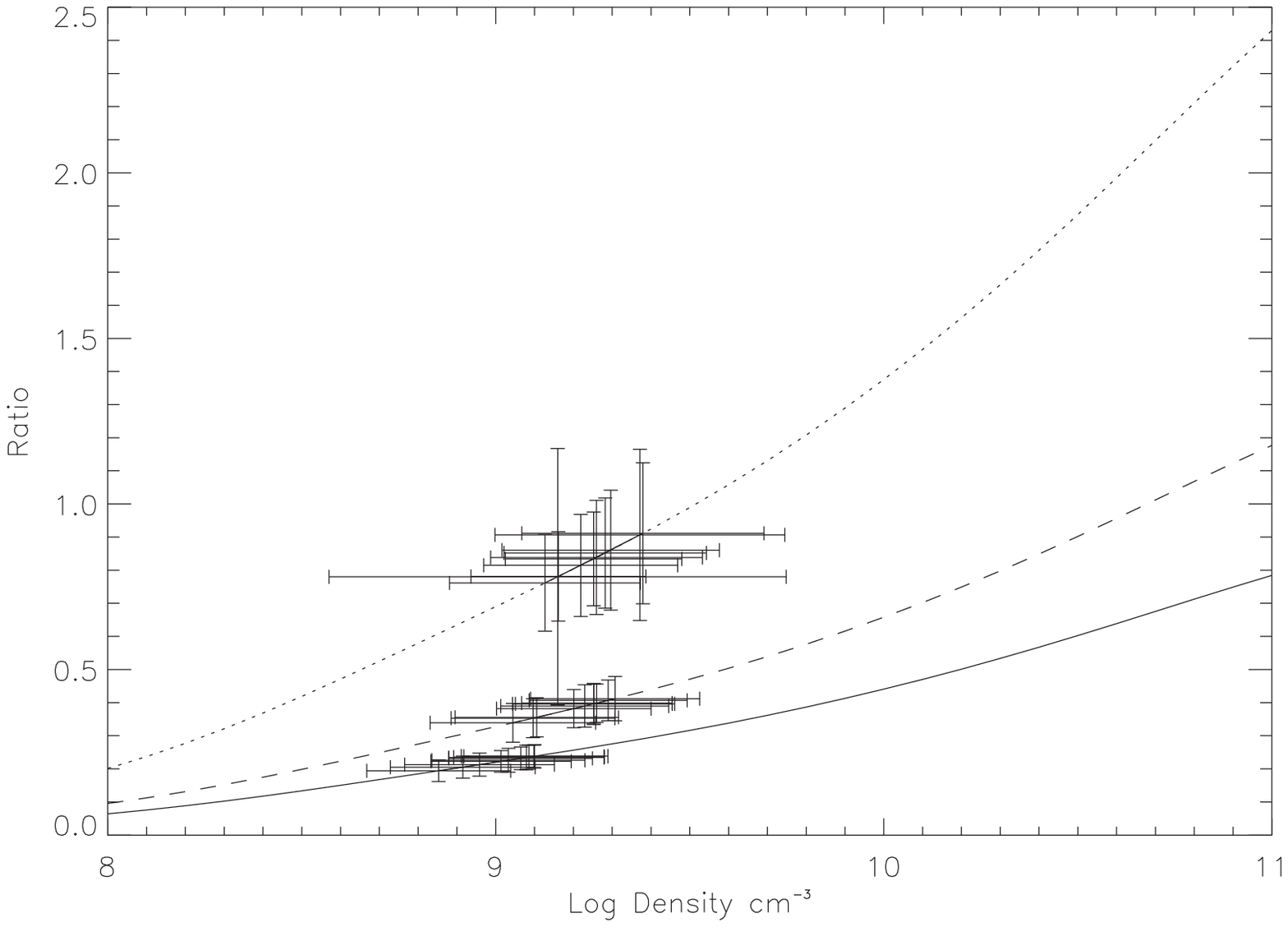}\plotone{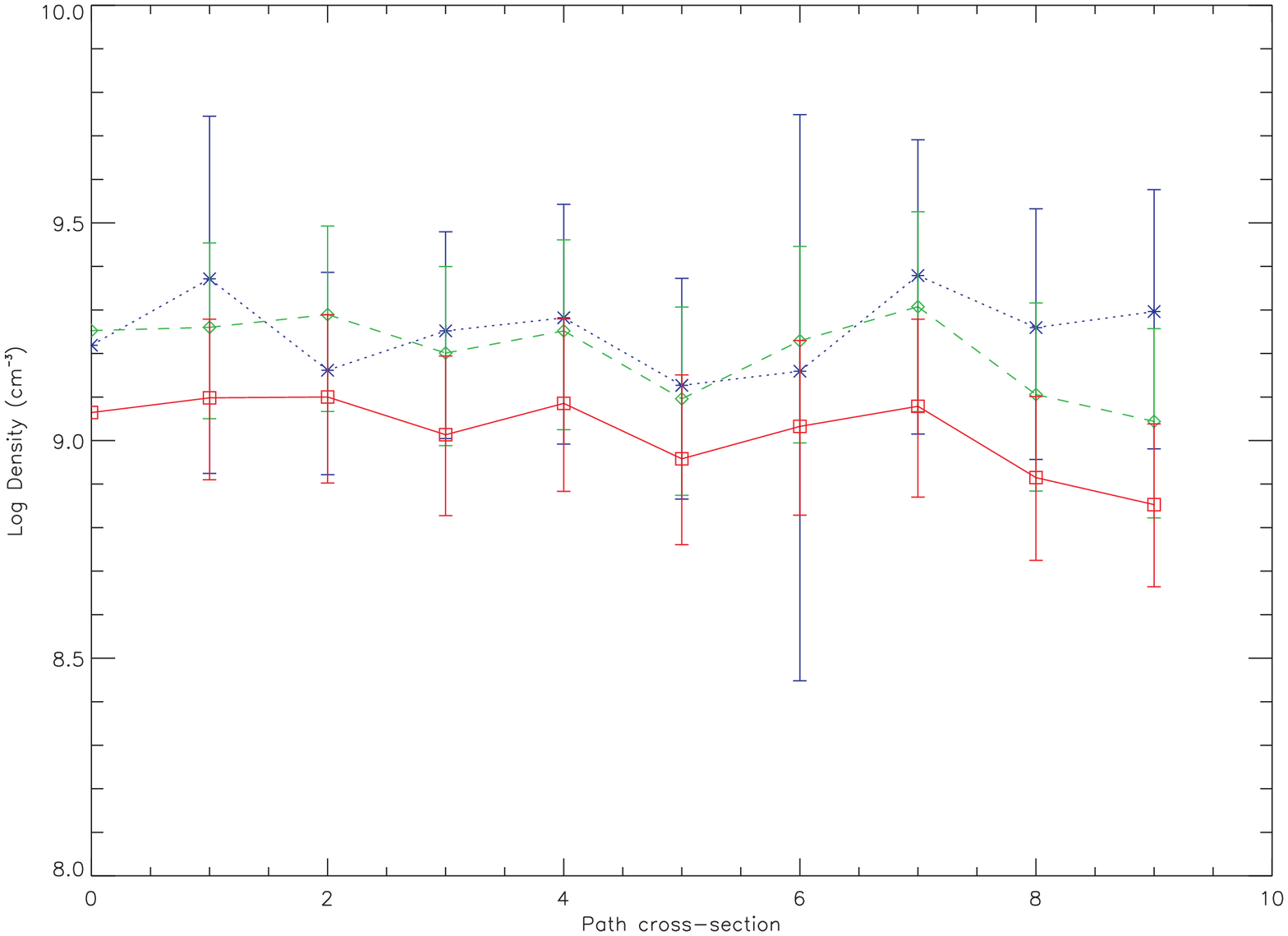}
\caption{a) Chianti curves of emission line ratio as a function of density (186.88/195.12)\mbox{\AA} (solid), (186.88/193.51)\mbox{\AA} (dashed), and (186.88/192.39)\mbox{\AA} (dotted), overplotted with the resulting data points from the loop path in the center panel of Figure~\ref{fig1}. b) Corresponding density profiles along the loop path where (186.88/195.12)\mbox{\AA} (solid red), (186.88/193.51)\mbox{\AA} (dashed green), and (186.88/192.39)\mbox{\AA} (dotted blue).}
\label{fig3}
\end{figure}

Figure \ref{fig3}a shows the CHIANTI density versus line ratio curves overplotted with the corresponding points along the loop path, indicating the good overlap between the densities obtained using the different ratio diagnostics. Figure~\ref{fig3}b shows the corresponding density profiles along the loop path,  demonstrating the good agreement between the density profiles obtained using the three iron density-sensitive ratios. The three profiles show a uniform density profile, consistent within the errors, along the base of the loop system where the wave is observed to propagate.

\subsection{Three-dimensional amplitude decay scale length}
\begin{deluxetable}{cc}
\tabletypesize{\scriptsize}
\tablecaption{Three-dimensional wave decay scale lengths\label{tab1}}
\tablewidth{0pt}
\tablehead{
\colhead{} & \colhead{Scale length $1/\alpha$ (km)} }
\startdata
STEREO A & $20,000^{+4000}_{-3000}$ \\
STEREO B &  $27,000^{+9000}_{-7000}$\\
 \tableline
\enddata
\end{deluxetable}

\begin{figure}[t]
\epsscale{0.9}
\centering
\rotate
\plotone{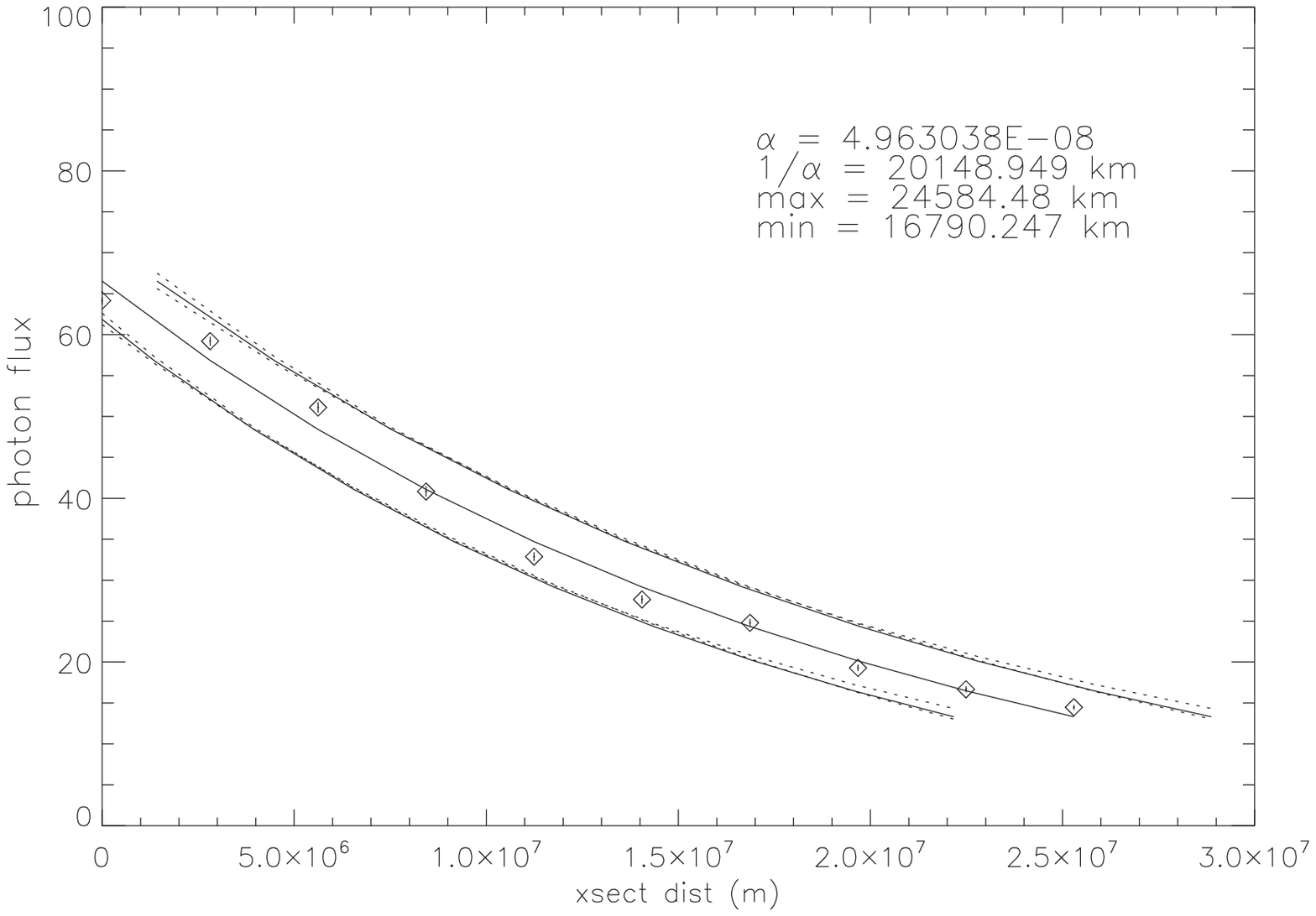}\plotone{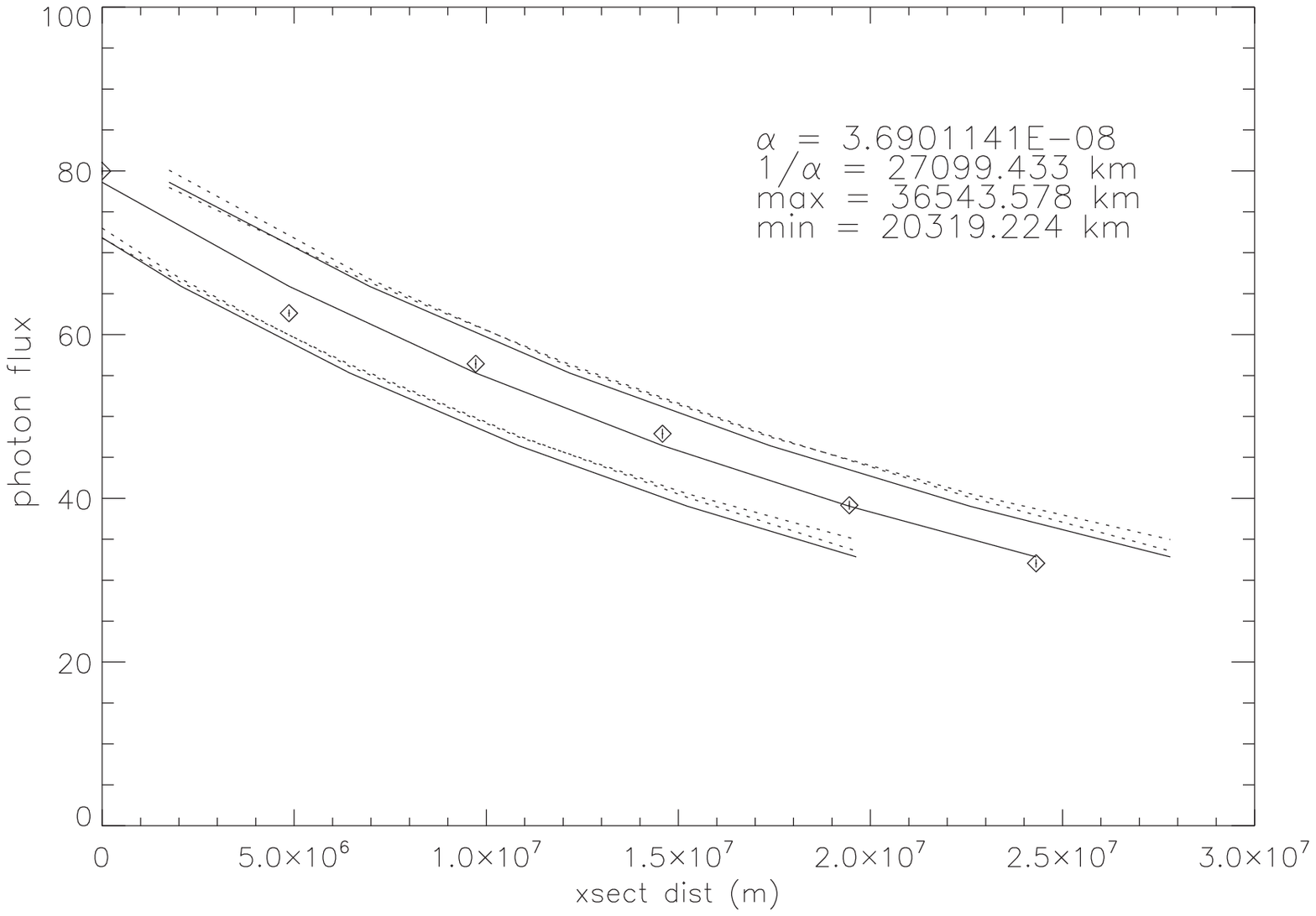}
\caption{a) The amplitude decay of the wave, along the rectified three-dimensional distance scale, measured from STEREO A (top) and STEREO B (bottom) with the fitted exponential decay curve (central solid curve), bounding curves indicating the error range due to the uncertainty in loop inclination (solid bounding curves), and error range with combined fitting errors (dotted bounding curves).  }
\label{fig4}
\end{figure}

Due to the projection effects of the loop inclination and the spacecraft line of sight, the measured amplitude profiles of the wave are projected onto the plane of the sky as viewed from each spacecraft.
The stereoscopic geometry results from \cite{me09a} are used to rectify the measured wave amplitude profiles to the true spatial scale, parallel to the wave propagation vector. 
It is assumed that the initial decay in both amplitude profiles corresponds to the same spatial location within the loop system. To ensure the same region of the loop is included within the fit to the amplitude decay, the same distance along the loop sysyem is fitted in the amplitude profiles on the A \& B rectified distance scales, shown in Figure~\ref{fig4}. 

Due to the uncertainty in the inclination of the loop, there is an uncertainty in rectified distance scale and a corresponding error on the decay length scale parameter of the fitted amplitude profiles, as plotted in Figure~\ref{fig4}. Also plotted are the max/min fitting errors on the max/min rectified distance scale uncertainties. The uncertainty curves in the plot indicate the max/min ranges on the distance scale uncertainty and fitting parameters. The maximum and minimum quoted scale lengths are the error ranges on the combined distance scale and max/min values of the fitted parameters. The fit to the amplitude decay gives the decay scale length of the wave, of the order 20,000~km, and associated uncertainties, given in Table~\ref{tab1}. The fitted length scales of the amplitude decay, observed from A \& B, agree considering these error ranges. 

\subsection{MHD Forward modelling}\label{forward_modelling}

\begin{figure}[h]
\epsscale{1.0}
\centering
\plotone{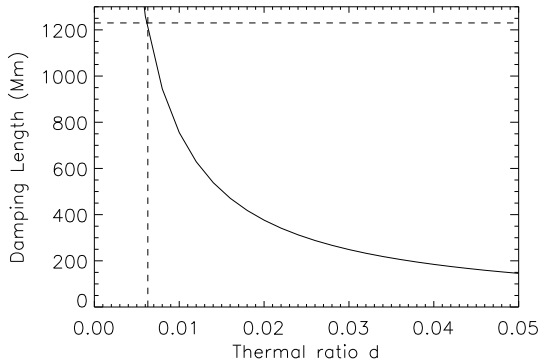}
\caption{The damping length in Mm as a function of the thermal conduction ratio $d$ as estimated from solving Equation~(\ref{eq:disp}). The dashed lines indicate $d=0.0063$ and the corresponding the damping length.}
\label{fig:thermcond}
\end{figure}

\begin{figure}[h]
\epsscale{1.0}
\centering
\rotate
\plotone{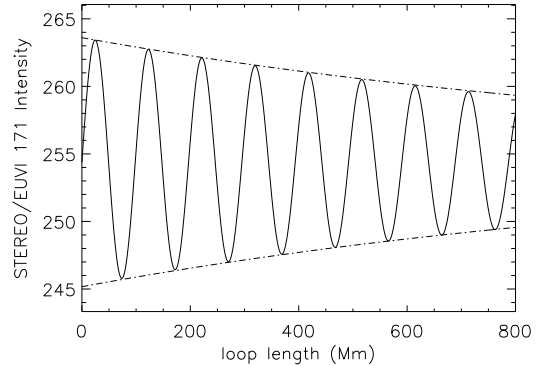}
\caption{Synthesised STEREO/EUVI emission as a function of length in Mm along the loop. The oscillatory signal is damped by optically thin radiation only. The dot-dashed lines correspond the exponential fits to the amplitude decay.}
\label{fig:radiation}
\end{figure}

\begin{figure}[h]
\epsscale{1.0}
\centering
\rotate
\plotone{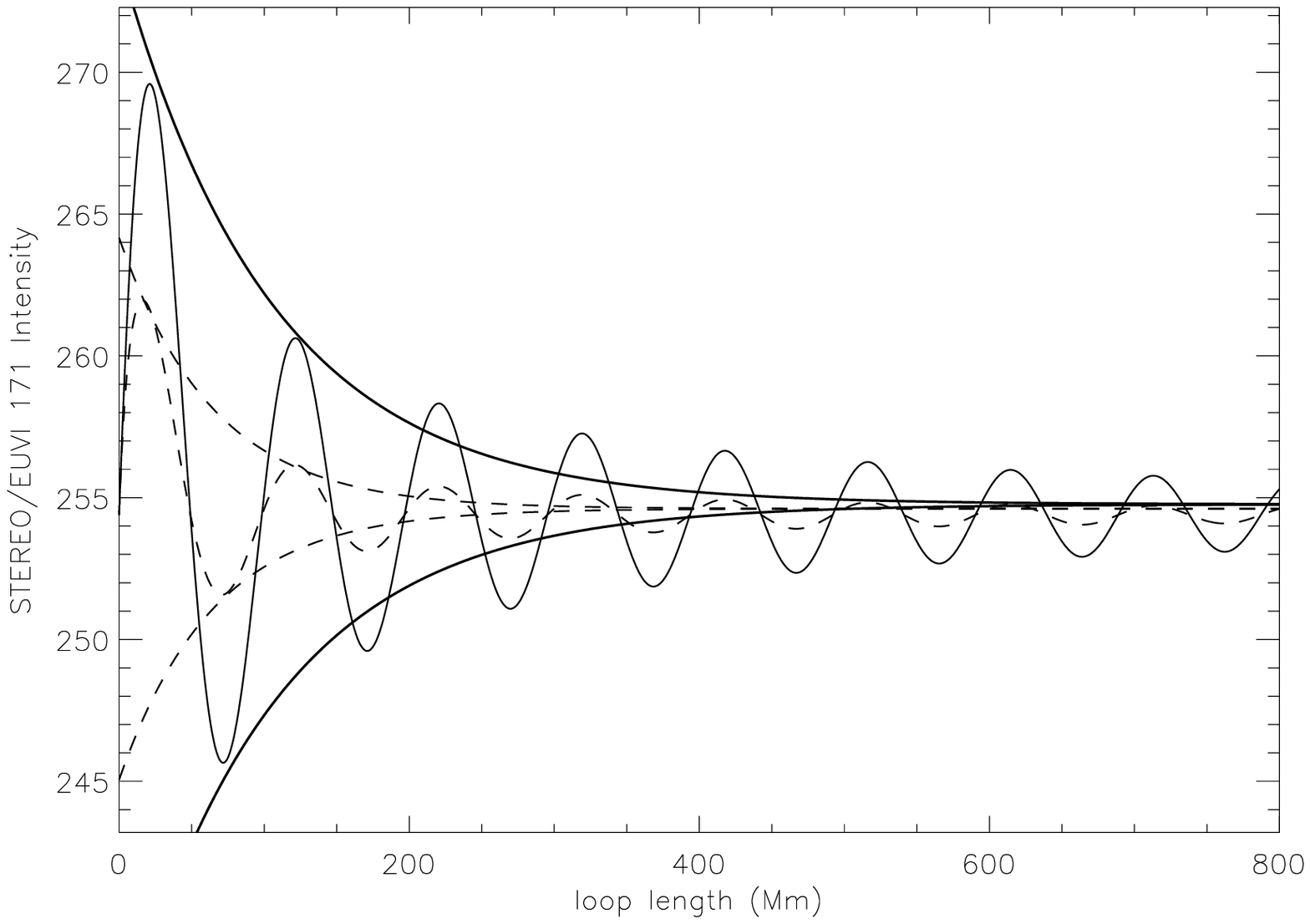}
\caption{Synthesised STEREO/EUVI A (solid lines) and B (dashed) emission as a function of length in Mm along the loop. The oscillatory signals are damped by area divergence only,  where viewed from STEREO A the loop has a basal radius $r_{0}=4.1$~Mm and loop expansion angle $\theta=5$ degrees, and viewed from STEREO B $r_{0}=5.3$~Mm and $\theta=27$ degrees. The enveloping lines correspond the exponential fits of the amplitude decay with damping lengths of 105~Mm for STEREO A, and 65~Mm for STEREO B.}
\label{fig:areadiv}
\end{figure}

In this section, we compare our observationally obtained damping length with the theoretical model of  \cite{dem&hoo03, dem&hoo04}, who considered damping by thermal conduction, optically thin radiation and area divergence. 

Using the 1D MHD model of  \cite{dem&hoo03, dem&hoo04}, we firstly calculate the value of the (dimensionless) thermal ratio $d$ and radiation ratio $r$ (Equations~(5) and (6) in \cite{dem&hoo04}, respectively). Using the observationally determined parameters of: loop background temperature of $T_0=0.83$ MK, background density $n_{e0}=10^{9.1}$ cm$^{-3}$, oscillation period $\tau=12$ minutes and standard coronal values for all other parameters, we find $d=0.0063$ and $r=0.13$. Note that these values are respectively a quarter (for the thermal conduction coefficient) and double (for the radiation coefficient) of the values used in the De Moortel \& Hood studies. Hence, for these specific parameters, the damping by thermal conduction will be reduced, whereas the effect of radiation will be enhanced, compared to the conclusions of  \cite{dem&hoo03, dem&hoo04}. 

\subsubsection{Thermal conduction}
Let us for the moment consider only thermal conduction (setting $r=0$). Assuming that all disturbances are expressed in terms of Fourier components, $\exp i (\omega t - k z)$, the following dispersion relation can be obtained
\begin{equation}
\omega^3 - i \omega^2 \gamma d k^2 - \omega k^2 + i d k^4=0
\label{eq:disp}
\end{equation}
(see Equation~(19) of  \cite{dem&hoo03}). As in \cite{dem&hoo03}, we can assume a fixed frequency $\omega$ and solve Equation~(\ref{eq:disp}) numerically, where the reciprocal of the imaginary part of $k$ corresponds to a damping length. This damping length has been plotted as a function of the thermal ratio $d$ in Figure~\ref{fig:thermcond}. (Note that in this figure, only the damping length for the slow mode is plotted, whereas the thermal mode has been omitted from this graph.) For the value of $d=0.0063$, we find a damping length of the order of 1230 Mm. Clearly, this value is far too large to be able to explain the observed damping lengths of the order of 20 Mm (see Section 4.2).

\subsubsection{Radiation}
Let us now turn to optically thin radiation. Here we can directly use the $\exp(-r z / \gamma)$ damping determined by \cite{dem&hoo04}, where $r=0.13$. (Note that the variable $z$ in this expression is a dimensionless quantity.) Using this value, an exponentially damped harmonic oscillation was constructed (using the observed values for temperature, density and the perturbation amplitude and period) and subsequently forward-modelled, using the response function of EUVI A, to obtain the synthesised emission as would be observed by STEREO/EUVI \citep[see][for details of the forward modelling procedure]{dem08}. The predicted intensity (in DN/s/pixel) is plotted in Figure~\ref{fig:radiation} as a function of loop length. Overplotted (dot-dashed line) is the exponential damping, using a damping length of the order of 1260 Mm. Similar to thermal conduction, optically thin radiation would not be able to account for the observed damping lengths. As the damping length obtained from thermal conduction and optically thin radiation are roughly similar, we could reasonably expect that the combined effect would give a damping length of about half of 1260 Mm, i.e. about 600 Mm. However, this is still at least an order of magnitude larger than the damping lengths estimated from the observations.

\subsubsection{Area divergence}
Finally, we look at the effect of area divergence. Assuming a general area divergence of the form $A(s) = \pi (r_0 + s \tan(\theta))^{2}$ (with $r_0$ the radius at the base of the loop, $\theta$ the expansion angle and $s$ the distance along the loop), leads to an amplitude decay of the form $A(s)^{-1/2}$ \cite[see][]{dem&hoo04}. Since the precise rotational geometry of the loop system is unknown, as an estimate of $r_0$ and $\theta$ we use the values from the bounding loop paths plotted in Fig.~\ref{fig1}, where from STEREO A  $r_{0}=4.1$~Mm and $\theta=5$ degrees and from STEREO B  $r_{0}=5.3$~Mm and $\theta=27$ degrees. The synthesised STEREO/EUVI emission of the resulting oscillatory signal ( i.e.~a harmonic oscillation with a period of 12 min and damping $\sim 1/\sqrt{A}$) is plotted in Figure~\ref{fig:areadiv}. To estimate a damping length, we have fitted an exponential decay to the first two peaks of the signal, as usually only these first two peaks are clearly identified in the observations. Using the measured values of $r_0$ and $\theta$, we obtain a damping length of 105~Mm for the STEREO A parameters, and a damping length of 65~Mm for STEREO B. Although still somewhat larger, this is now of a similar order of magnitude as the observed damping length, especially keeping in mind that thermal conduction and optically thin radiation will cause some additional damping.

\subsection{Comparison to dissipation theory}
As referenced in the introduction to this paper, there has been much work on the theoretical modelling of slow mode propagation and dissipation within coronal loops investigating effects on the wave amplitude such as gravitational stratification and magnetic field divergence, along with dissipative mechanisms including thermal conduction, compressive viscosity, and optically thin radiation. 

\cite{ofm02b} analyzed 35 cases of damped Doppler observations in hot $>$6~MK coronal loops using the SOHO/SUMER instrument, interpreted as signatures of standing slow magnetoacoustic modes \citep{wan02, wan03}. Using a 1D MHD model incorporating thermal conduction, viscosity and radiation, they found that the observed rapid damping of the 5-30 minute period oscillations can be explained by thermal conduction. 

\cite{dem&hoo03} study a one dimensional model including the effects of thermal conduction and compressive viscosity on 5-minute period waves in 1~MK loops. They find that thermal conduction is dominant and that the slow mode has a minimum damping length due to thermal conduction alone.
\cite{dem&hoo04} extend this model further including the effects of magnetic field divergence and gravitational stratification, and find that thermal conduction and field divergence are the dominant factors. 

\cite{kli04} investigate analytical and numerical one dimensional models of 5-minute waves in $\sim$1~MK loops, and find significant damping due to thermal conduction and the observing instrument response function. The simulations are consistent with damping due to classical thermal conduction alone without additional damping mechanisms, or unusually large conduction coefficient. They investigate some previous observations from \cite{dem02a} with an estimated intensity scale length based on their defined detection length. The advantage of the observations that we present here are that they are essentially observations of a monochromatic wave. Thus allowing a direct measurement of the intensity scale length with the geometrical projection effects removed using stereoscopic observations. 
 
Forward modelling techniques have been applied to slow mode propagation \citep[see][]{dem08, owe09}.  \cite{dem08} show that observational parameters such as phase, damping rate, and period can be affected by plasma emission processes. i.e. the shape of the ionisation balance for a particular line can lead to a reduction in intensity with an increased temperature, depending on the value of the background temperature of the plasma and the amplitude of the temperature perturbations. The loop system analyzed here was found to have a uniform temperature profile at the base of the loops \citep{me09b}, and, consistent with this, the observed waves also have a constant phase speed \citep{me09a}. The observed values of the loop temperature and density, along with the parameters of the observed wave propagation, were incorporated into the forward modelling described in Section~\ref{forward_modelling}. 

The previous studies investigating slow magnetoacoustic waves in coronal loops of temperatures around 1~MK and higher, described above, find thermal conduction is the dominant damping mechanism. Thermal conduction is a valid explanation of the observed damping, even for the observed low oscillation frequencies, in the type of $\sim$6~MK loops observed by SUMER \citep{ofm02b}; this is also the case for 1~MK loops with oscillation periods around 5-minutes \citep{dem&hoo03, kli04}.  However, the implications for the class of sub-million degree ``cool loops'' with long oscillation periods ($\sim$10-minutes) observed here has not previously been considered. In the case of the temperature and density of the loop observed here, we find that thermal conduction alone is insufficient to account for the observed damping of the wave, since the relatively cool loop temperature and long oscillation period lead to very weak thermal conduction; therefore, area divergence has the dominant effect on the observed decay of the wave.
It should be noted that magnetic field divergence may explain the observed decay of the wave amplitude, however, this is a geometrical effect and not a dissipation mechanism of the wave energy. 

\section{Conclusions}
The results presented here determine the true, three-dimensional, dissipation length scale of a slow magnetoacoustic mode in the solar corona. The decay scale length at the maximum amplitude of the wave  observed by STEREO A \& B is $20,000^{+4000}_{-3000}$km and $27,000^{+9000}_{-7000}$km respectively. The agreement between the results from both spacecraft, considering the uncertainties, places a quantitive constraint on theoretical models of slow mode propagation and dissipation mechanisms within the corona. 

Using the spectroscopic diagnostics of Hinode/EIS, the density profile of the loop system is found to be uniform where the wave propagates. This coincides with a uniform temperature profile, measured by \cite{me09b}. The three different density diagnostic ratios, under test here, give consistent density profiles within the uncertainties. This suggests that they are are suitable diagnostics for these types of loop systems, although, the (186.88/195.12)\mbox{\AA} yields more precise results due to greater signal to noise. The agreement also implies that the assumptions regarding negligible blend contributions to the \ion{Fe}{12} 195.12\mbox{\AA} line, described in Section~\ref{loop_density_analysis}, are valid. 

The MHD modelling and simulated synthetic observations presented here show that thermal conduction is insufficient to explain the observed short decay length of the waves observed with STEREO/EUVI. In contrast to previous studies of slow mode damping in coronal loops which can be explained by thermal conduction, in the case presented here, the divergence of the magnetic field is found to be most significant, due to the temperature, density, and oscillation period within the observed loop, 
damping due to thermal conduction is decreased and damping due to radiation is enhanced to be of the same order.
Field line divergence alone can account for the observed damping to within a factor of three. To determine if the combined effects of radiation, thermal conduction, and field line divergence can fully account for the observed damping length a full numerical study incorporating all these effects simultaneously will be required in future.

\acknowledgments
This research is supported by the Science and Technology Facilities Council (STFC) under grant number ST/F002769/1. IDM acknowledges support from a Royal Society University Research Fellowship. Hinode is a Japanese mission developed and launched by ISAS/JAXA, with NAOJ as domestic partner and NASA and STFC (UK) as international partners. It is operated by these agencies in co-operation with ESA and NSC (Norway). 
The STEREO/SECCHI data used here are produced by an international consortium of the Naval Research Laboratory (USA), Lockheed Martin Solar and Astrophysics Lab (USA), NASA Goddard Space Flight Center (USA) Rutherford Appleton Laboratory (UK), University of Birmingham (UK), Max-Planck-Institut f\"{u}r Sonnensystemforschung(Germany), Centre Spatiale de Liege (Belgium), Institut d'Optique Théorique et Applique\'{e} (France), Institut d'Astrophysique Spatiale (France).
CHIANTI is a collaborative project involving researchers at NRL (USA) RAL (UK), and the Universities of: Cambridge (UK), George Mason (USA), and Florence (Italy).
M.S. Marsh would like to acknowledge the encouragement of L.E. Marsh.

{\it Facilities:} \facility{Hinode (EIS)} \facility{STEREO (EUVI)}.

\bibliographystyle{apj}
\bibliography{ms}

\end{document}